\documentclass[aps,pra,twocolumn,showpacs,floatfix,superscriptaddress]{revtex4-1}
\usepackage{graphicx}
\usepackage{bm}
\usepackage{amsmath}

\begin{document}

\title{Interferometric distillation and determination of
unknown two-qubit entanglement}
\author{S.-S. B. Lee}
\author{H.-S. Sim}
\email{hssim@kaist.ac.kr}
\affiliation{Department of Physics, Korea Advanced Institute of Science and Technology,
Daejeon 305-701, Korea}
\begin{abstract}
We propose a scheme for both distilling and quantifying
entanglement, applicable to individual copies of an arbitrary
unknown two-qubit state. It is realized in a usual two-qubit
interferometry with local filtering.
Proper filtering operation for the maximal distillation of the state is achieved,
by erasing single-qubit interference,
and then the concurrence of the state is
determined directly from the visibilities of two-qubit interference.
We compare the scheme with full state tomography.
\end{abstract}

\maketitle

{\it Introduction.}---
Multiparticle interference is a striking phenomenon connecting
with quantum entanglement.
For pure states, the connection is rather
straightforward. In a two-particle interferometry
\cite{Ghosh,Horne}, the interference visibility gives the
concurrence \cite{Bennet_conc,Wootters_conc}, a widely used
entanglement measure, of the two-particle pure state \cite{Jakob}.
In a multiparticle Aharonov-Bohm interferometry \cite{Sim}, the
visibility can be used to prove the quantum nonlocality of
Greenberger-Horne-Zeilinger entanglement \cite{Greenberger}. For
mixed states, however,
multiparticle interference comes from a mixture of
entanglement and classical correlation, and it is hard to distinguish
the two different correlations.
It is interesting to find a way to extract
entanglement from the interference, which is the aim of this work.

In quantum information research, there are strong demands of
distilling and quantifying entanglement \cite{Horodecki_rev}.
Currently available schemes are of two types, one using
multiple copies of a target state and the other using individual
copies. Since the multiple copies are harder to prepare in
laboratory in general,
it may be necessary to explore further the latter type.
The distillation of the latter type has been done using local filtering,
for a known two-qubit state \cite{Kwiat} or after full state tomography \cite{Wang}.
And no scheme of the latter type has been proposed
for directly quantifying an entanglement measure,
such as concurrence, of an arbitrary mixed state in experiments;
note that the existing schemes of the former type for determining concurrence
have not been realized \cite{Horodecki_copy} or provide
a lower bound of concurrence \cite{Mintert_mixed2} for mixed states, while concurrence was
recently determined in experiments by using two copies of a pure state \cite{Walborn}.
Therefore, it is valuable to find a scheme of the latter type for distilling and
directly determining entanglement
of an unknown state (without full state reconstruction).

In this work, we propose an interferometric scheme for both
distilling and determining entanglement, applicable to individual
copies of an {\em arbitrary unknown} two-qubit state.
It can be realized in a two-qubit interferometry with local filtering \cite{Kwiat,Wang}.
The maximal distillation (the normal form \cite{Verstraete_filter}) of
the state is first achieved, by iteratively erasing single-qubit interference,
and then the concurrences of both the initial and
the distilled states are {\em directly} determined from the visibilities of two-qubit interference.
This quantification is based on our important findings that the two-qubit interference
shows {\em three} different ``local'' extrema (visibilities) in general and that
when the single-qubit interference is fully erased, the three extrema give the Lorentz singular
values \cite{Verstraete_filter},
a linear combination of which gives the concurrences.
Our scheme is conceptually different from full state tomography
and practically useful.

\begin{figure}[bt]
\includegraphics[width=0.4\textwidth]{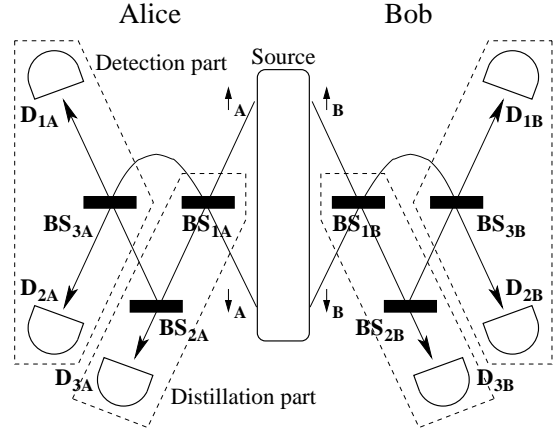}
\caption{{ Two-qubit interferometry with local filtering.} It
consists of a source, distillation parts, and detection parts.
The source generates individual copies of an arbitrary unknown state of
two qubits A and B, each represented by pseudospins $\uparrow_j$ and
$\downarrow_j$, $j=\mathrm{A},\mathrm{B}$.  Qubit $\mathrm{A}$
flies to the detectors ($\mathrm{D}_\mathrm{A}$) of Alice, passing
through three beam splitters (BS), while $\mathrm{B}$ to Bob
($\mathrm{D}_\mathrm{B}$); see solid arrows. The state is
transformed into its maximally distilled state in the distillation parts,
and then its concurrence is determined
by measuring the visibilities of two-qubit interference in
the detection parts.
} \label{setup}
\end{figure}

{\it Two-qubit interferometry with local filtering.}--- We introduce
the interferometry (Fig.~\ref{setup}). The source generates
individual copies of a state of two qubits A and B, which fly to
Alice and Bob, respectively. The qubit degree of freedom, the
pseudospin, can be photon polarization, particle path,
particle spin, etc. For illustration, we choose particle path
as the pseudospin, by considering two particles A and B, each
injected to either the upper (pseudospin up) or the lower path
(down) of its side.

The $4 \times 4$ density matrix $\hat{\rho}$ of the initial
two-qubit state is represented by using pseudospin basis states, and
written by using a real matrix $\textbf{R}$ as \cite{Schlienz}
\begin{equation}
\hat{\rho} = \frac{1}{4} \sum_{l,l'=0,1,2,3} R_{l l'} \hat{\sigma}_l
\otimes \hat{\sigma}_{l'}, \,\,\,\,\, R_{l l'} \equiv \mathrm{Tr}
(\hat{\rho} \hat{\sigma}_l \otimes \hat{\sigma}_{l'}),
\label{R_MATRIX}
\end{equation}
where $\mathrm{Tr} (\cdots)$ means the trace of matrix, 
$\hat{\sigma}_0$ is the $2 \times 2$ identity matrix, and
$\hat{\sigma}_l$'s ($l=1,2,3$) are the Pauli matrices. Then the
single-qubit states ($\hat{\rho}_j \equiv \mathrm{Tr}_{\bar{j}}
\hat{\rho}$) of Alice and Bob
are represented by 4-vectors $(R_{00},R_{10},R_{20},R_{30})$ and
$(R_{00},R_{01},R_{02},R_{03})$, respectively;
$\mathrm{Tr}_{\bar{j}}(\cdots)$ means the trace over the degrees of
freedom of qubit $\bar{j}$ $(\ne j)$.

In the distillation parts, which are absent in usual
interferometries \cite{Ghosh,Horne}, Alice (Bob) has local operation
$\hat{D}_j (f_j, \theta_{\textrm{dis},j}, \phi_{\textrm{dis},j})
\equiv \hat{F}(f_j) \hat{U}(\theta_{\textrm{dis},j},
\phi_{\textrm{dis},j})$ on qubit $j = \mathrm{A}$ ($\mathrm{B}$). It
transforms the initial state $\hat{\rho}$ into its normal form
\cite{Verstraete_filter,Verstraete_normal} $\hat{\rho}_{\rm dis}$,
its maximally distilled state \cite{Kent},
\begin{equation}
\hat{\rho}_\mathrm{dis} = (\hat{D}_\mathrm{A} \otimes
\hat{D}_\mathrm{B}) \hat{\rho} (\hat{D}_\mathrm{A} \otimes
\hat{D}_\mathrm{B})^\dagger. \label{DISTILL}
\end{equation}
The local filtering $\hat{F}(f_j)$ and the rotation
$\hat{U}(\theta_{\textrm{dis},j}, \phi_{\textrm{dis},j})$ of qubit
$j$ are supported by two beam splitters, $\mathrm{BS}_{2j}$ and
$\mathrm{BS}_{1j}$, respectively, and represented as
\begin{equation}\label{filtering}
\hat{F} (f) = \begin{pmatrix}1 & 0 \\
0 & f \\
\end{pmatrix}, \,\,\,
\hat{U} (\theta,\phi)= \begin{pmatrix}
\cos \frac{\theta}{2} & \sin \frac{\theta}{2} e^{- i \phi} \\
- \sin \frac{\theta}{2} e^{i \phi} & \cos \frac{\theta}{2} \\
\end{pmatrix}.
\end{equation}
Here, $0 \le f_j \le 1$ is the filtering parameter controlled by the
reflection amplitude of $\mathrm{BS}_{2j}$, $\theta_{\textrm{dis},j}
\in [0, \pi]$ parameterizes the transmission at $\mathrm{BS}_{1j}$,
and $\phi_{\textrm{dis},j} \in [0, 2\pi]$ is the phase shift. In the
filtering operation $\hat{F}(f_j)$, particle $j$ is abandoned with
probability $1-f_j^2$ when it flies along the lower path after
scattering by $\mathrm{BS}_{1j}$.
Whether qubit $j$ is filtered off or not
is certified at detector $\mathrm{D}_{3j}$. The two beam splitters
constitute the minimal setup for the distillation. This is
understood from the fact \cite{Verstraete_filter} that each local
operation on qubit $j$ corresponds to a Lorentz transformation
of the 4-vector of qubit $j$. $\hat{F}$ and
$\hat{U}$ correspond to the Lorentz boost and the spatial rotation,
respectively. We emphasize that the Lorentz boost mathematically
introduced in Ref. \cite{Verstraete_filter}
is physically realized here by
$\hat{F}$, using beam splitter $\mathrm{BS}_{2j}$. We will see later
how
$\hat{D}_j$ is efficiently found for an unknown state $\hat{\rho}$.

In the detection parts, Alice and Bob count the number $n_{ij}$ of
the particles $j$ arriving at detector $\mathrm{D}_{ij}$, $i=1,2$,
during measurement time, long enough to get the statistical average
of coincidence correlation
$\langle n_{i \mathrm{A}} n_{i' \mathrm{B}} \rangle$ for a given setting of all
the beam splitters. They tune $\mathrm{BS}_{3 \mathrm{A}}$ and
$\mathrm{BS}_{3 \mathrm{B}}$ to see single- and two-qubit
interferences 
in $\langle n_{1j} (n_{1 \bar{j}} + n_{2 \bar{j}})
\rangle$ and $\langle \delta n_{1 \mathrm{A}} \delta n_{1
\mathrm{B}} \rangle \equiv \langle n_{1 \mathrm{A}} n_{1 \mathrm{B}}
\rangle - \langle n_{1 \mathrm{A}} (n_{1 \mathrm{B}} + n_{2
\mathrm{B}}) \rangle \langle n_{1 \mathrm{B}} (n_{1 \mathrm{A}} +
n_{2 \mathrm{A}}) \rangle$, respectively;
the other correlations involving $\mathrm{D}_{2j}$ contain the same information.
Here, the
number $n_{ij}$ is normalized by the total number $N$ of states
ending in neither $\mathrm{D}_{3 \mathrm{A}}$ nor $\mathrm{D}_{3
\mathrm{B}}$, and $\langle n_{ij} (n_{1 \bar{j}} + n_{2 \bar{j}})
\rangle$ means the average number $n_{ij}$ of qubit $j$ in the
events where the other qubit $\bar{j}$ $(\ne j)$ is not filtered
(does not end in $\mathrm{D}_{3 \bar{j}}$). The qubit rotation at $\mathrm{BS}_{3j}$
is represented by $\hat{U} (\theta_{\textrm{det},j},
\phi_{\textrm{det},j})$, where $\theta_{\textrm{det},j} \in [0,
\pi]$ and $\phi_{\textrm{det},j} \in [0, 2 \pi]$.
Note that the
phase accumulation of particle $j$ along its path
is absorbed in the rotation angles $\phi_{{\rm
dis},j}$ and $\phi_{{\rm det}, j}$.

The visibilities of single- and two-qubit interferences are defined,
respectively, as \cite{Jaeger}
\begin{equation}\label{VISIBILITY}
\begin{split}
\mathcal{V}_{j = \mathrm{A}, \mathrm{B}} &= W[\langle n_{1j} (n_{1
\bar{j}} + n_{2 \bar{j}})
\rangle], \\
\mathcal{V}_\mathrm{AB} &= W[\langle \delta n_{1 \mathrm{A}} \delta
n_{1 \mathrm{B}} \rangle + 1/4],
\end{split}
\end{equation}
where $W[x] \equiv (\mathrm{max} [ x ] - \mathrm{min}
[ x ] ) / (\mathrm{max} [ x ] + \mathrm{min} [ x ] )$ and
$\mathrm{max} [ x ]$ ($\mathrm{min} [ x ]$) means the local maxima
(minima) of $x$ over the parameter space of $\mathrm{BS}_{3
\mathrm{A}}$ and $\mathrm{BS}_{3 \mathrm{B}}$, $\{
\theta_{\textrm{det}, \textrm{A}}, \phi_{\textrm{det}, \textrm{A}},
\theta_{\textrm{det}, \textrm{B}}, \phi_{\textrm{det}, \textrm{B}}
\}$. Since the mean value of $\langle \delta n_{1 \mathrm{A}} \delta
n_{1 \mathrm{B}} \rangle$ over the space is zero, the {\it ad hoc}
factor $1/4$ is added in Eq.~\eqref{VISIBILITY}
so that the values of
$\mathcal{V}_\mathrm{AB}$ are equal to the local maxima of
$\langle \delta n_{1 \mathrm{A}} \delta n_{1 \mathrm{B}} \rangle$.
Note that one needs to tune the transmission probability $\cos^2
\frac{\theta_{\textrm{det}, j}}{2}$ of $\mathrm{BS}_{3j}$, to obtain
the information of the diagonal parts of $\hat{\rho}$ \cite{Jaeger}.

{\it Entanglement distillation.}--- We first explain how to
transform $\hat{\rho}$ into $\hat{\rho}_\mathrm{dis}$. It is based
on the facts \cite{Verstraete_normal} that
local density matrices $\hat{\rho}_{\mathrm{dis},j} \equiv
\mathrm{Tr}_{\bar{j}} \hat{\rho}_\textrm{dis}$ of the normal form
are proportional to the identity matrix
and that any initial state $\hat{\rho}$ can be transformed
iteratively to its normal form by filtering operations.
In our interferometry, we find that these facts are realized as follows.
When
$\hat{\rho}_\mathrm{dis}$ is achieved in the distillation parts,
the single-qubit interference visibilities
$\mathcal{V}_\mathrm{A}$ and $\mathcal{V}_\mathrm{B}$ vanish, since
the identity generates no interference.
Thus, one can
obtain $\hat{D}_j$ in an iterative way such that Alice and Bob
alternately tune her/his beam splitters of the distillation parts to
make the visibility $\mathcal{V}_j$ of her/his single-qubit
interference vanish, until $\mathcal{V}_\mathrm{A}$ and
$\mathcal{V}_\mathrm{B}$ both vanish simultaneously.

We describe each iteration step. In the $(2k-1)$-th step,
$k=1,2,\cdots$, Alice observes the single-qubit interference signal
in $\langle n_{1 \mathrm{A}} (n_{1 \mathrm{B}} + n_{2 \mathrm{B}})
\rangle$ by tuning $\mathrm{BS}_{3 \mathrm{A}}$, with setting her
parameters as $(f_\mathrm{A}, \theta_{\mathrm{dis},\mathrm{A}}) =
(1,0)$ but without tuning $(f^{(2k-2)}_\mathrm{B},
\theta^{(2k-2)}_{\mathrm{dis},\mathrm{B}},
\phi^{(2k-2)}_{\mathrm{dis},\mathrm{B}})$ fixed by Bob; in the first
step, Bob starts with $(f^{(0)}_\mathrm{B},
\theta^{(0)}_{\mathrm{dis},\mathrm{B}}) = (1,0)$. By comparing the
signal with its general form, $\langle n_{1 \mathrm{A}} (n_{1
\mathrm{B}} + n_{2 \mathrm{B}}) \rangle = (1 +
\vec{\gamma}_\mathrm{A} \cdot \vec{v}_\mathrm{A})/2$, Alice
determines $\vec{\gamma}_\mathrm{A} \equiv (\gamma_\mathrm{A1},
\gamma_\mathrm{A2}, \gamma_\mathrm{A3})$, which in fact represents
the spatial part of the 4-vector of qubit $\mathrm{A}$; the general
form has only a pair of extrema $\pm |\vec\gamma_\mathrm{A}|$,
i.e., 
$\mathcal{V}_\mathrm{A}$ $(= |\vec\gamma_\mathrm{A}|)$ is single-valued.
Here $\vec{v}_j \equiv (\sin \theta_{\mathrm{det},j} \cos
\phi_{\mathrm{det},j}, \sin \theta_{\mathrm{det},j} \sin
\phi_{\mathrm{det},j}, \cos \theta_{\mathrm{det},j})$ is the
rotation vector of $\mathrm{BS}_{3j}$. Then, by setting
$f_\mathrm{A}^{(2k-1)} = \sqrt{(1 -
|\vec{\gamma}_\mathrm{A}|)/(1 + |\vec{\gamma}_\mathrm{A}|)}$
and the rotation vector of $\mathrm{BS}_{\mathrm{1A}}$ in the $(2k-1)$-th step as
$\vec{v}_{\mathrm{dis,A}}^{(2k-1)} = - \vec{\gamma}_\mathrm{A} / |\vec{\gamma}_\mathrm{A}|$,
Alice
achieves the situation that $\mathcal{V}^{(2k-1)}_\mathrm{A}$
vanishes; after the setting, $\vec{\gamma}_\mathrm{A}$ rotates to
be parallel to $\vec{z} = (0,0,-1)$ at $\mathrm{BS}_\mathrm{1A}$, and
then vanishes by the filtering at $\mathrm{BS}_\mathrm{2A}$.
Next, Bob performs his $(2k)$-th step in the same way as Alice's
$(2k-1)$-th step, except for the exchange $\mathrm{A}
\leftrightarrow \mathrm{B}$. After such iterations, the distillation
parameters converge to $(f_j, \theta_{\mathrm{dis},j},
\phi_{\mathrm{dis},j})$, at which $\mathcal{V}_{\mathrm{A}} = \mathcal{V}_{\mathrm{B}} = 0$
and $\hat{\rho}_\mathrm{dis}$ is obtained.

{\it Entanglement determination}.--- Before discussing
entanglement quantification, we first show
an interesting feature of $\mathcal{V}_\mathrm{AB}$.
For a state $\hat{\rho}'$,
transformed from $\hat{\rho}$ by an arbitrary setting of the
distillation parts, we derive a compact form of the cross-correlation,
$\langle \delta n_{1 \mathrm{A}} \delta n_{1 \mathrm{B}} \rangle
= \frac{1}{4} \vec{v}_\mathrm{A} \mathbf{Q}' \vec{v}_\mathrm{B}^T$,
where $\vec{v}_j$ is the rotation vector of $\mathrm{BS}_{3j}$,
the column vector $\vec{v}^T_\mathrm{B}$ is the transpose of
$\vec{v}_\mathrm{B}$, $\textbf{Q}'$ is the $3 \times 3$ matrix
defined by $Q_{l l'}' = R'_{l l'}/R'_{00} - R'_{l 0} R'_{0 l'} /
(R'_{00})^2$, $l,l' = 1,2,3$, and $\textbf{R}'$ is the real
parametrization of $\hat{\rho}'$ in Eq. (\ref{R_MATRIX}); the number
normalization by $N$ gives the factors $1/R'_{00}$ and
$1/(R'_{00})^2$ in $Q_{l l'}'$.
From this compact form and the fact that $\vec{v}_{j =
\mathrm{A},\mathrm{B}}$ spans over the surface of unit sphere,
it is easy to see that
$\langle \delta n_{1 \mathrm{A}} \delta n_{1 \mathrm{B}} \rangle$
has {\em three} pairs of ``local'' extrema $\pm \lambda_l$'s ($\lambda_1
\ge \lambda_2 \ge \lambda_3 \ge 0$),
i.e., $\mathcal{V}_\mathrm{AB}$
has the three values $\lambda_l$'s, and that
$\lambda_l$'s are
identical to the {\em singular values} of $\textbf{Q}'$ up to sign
factor. Here, $\lambda_1$ is the global maximum of
$\langle \delta n_{1 \mathrm{A}} \delta n_{1 \mathrm{B}} \rangle$,
while $\lambda_2$ ($\lambda_3$) is the maximum over the space
of $\vec{v}_{\mathrm{A}}$ and $\vec{v}_{\mathrm{B}}$ orthogonal to
$\vec{v}_{\mathrm{A},1}$ and $\vec{v}_{\mathrm{B},1}$
($\vec{v}_{\mathrm{A},1}$, $\vec{v}_{\mathrm{B},1}$,
$\vec{v}_{\mathrm{A},2}$, and $\vec{v}_{\mathrm{B},2}$),
where $\vec{v}_{j,l}$ is
the rotation vector of $\mathrm{BS}_{3j}$
at which $\langle \delta n_{1 \mathrm{A}} \delta
n_{1 \mathrm{B}} \rangle$ shows
$\lambda_l$.

The above finding becomes very useful when
$\hat{\rho}_\mathrm{dis}$ is achieved in the
distillation parts ($\hat{\rho}' = \hat{\rho}_\mathrm{dis}$ and
$\textbf{R}' = \textbf{R}_\mathrm{dis}$). In this case,
the visibilities $\lambda_l$'s give
the Lorentz singular values \cite{Verstraete_normal} of
the initial state as
\begin{equation}
s_0 = \frac{N}{M f_\mathrm{A} f_\mathrm{B}}, \,\,\, s_1 = s_0
\lambda_1, \,\,\, s_2 = s_0 \lambda_2, \,\,\, s_3 = q s_0 \lambda_3,
\label{SINGULAR}
\end{equation}
since they are equal to the singular values of
$\textbf{R}_\mathrm{dis}/(f_\mathrm{A} f_\mathrm{B})$,
$R_{\mathrm{dis},l0} = R_{\mathrm{dis},0l} = 0$, and
$R_{\textrm{dis},00} = N / M$. Here,
$M$ is
the total number of injection of
$\hat{\rho}$ from the source, $q =
\mathrm{Det}(\vec{v}_{\mathrm{A},1}, \vec{v}_{\mathrm{A},2},
\vec{v}_{\mathrm{A},3}) \mathrm{Det}(\vec{v}_{\mathrm{B},1},
\vec{v}_{\mathrm{B},2}, \vec{v}_{\mathrm{B},3})$ is the sign factor
guaranteeing the correct singular value decomposition,
$\mathrm{Det}(\cdots)$ means matrix determinant,
and $(\vec{v}_{j,1}, \vec{v}_{j,2}, \vec{v}_{j,3})$ is
the matrix whose columns are $\vec{v}_{j,l}$'s. Using the relation
\cite{Verstraete_filter} between concurrence
and Lorentz singular values, we find an important result that the concurrences
$\mathcal{C}$ of $\hat{\rho}$ and $\hat{\rho}_\mathrm{dis}$ are
{\em directly} obtained from $\mathcal{V}_\mathrm{AB}$,
\begin{equation}\label{Concurrence1}
\begin{split}
\mathcal{C} (\hat{\rho}) &= s_0 \mathcal{C} (\hat{\rho}_\mathrm{dis}),\\
\mathcal{C} (\hat{\rho}_\mathrm{dis}) &= \mathrm{max} [0,
\frac{1}{2} (-1 + \lambda_1 + \lambda_2 - q \lambda_3)].
\end{split}
\end{equation}

\begin{figure}
\includegraphics[width=0.40\textwidth]{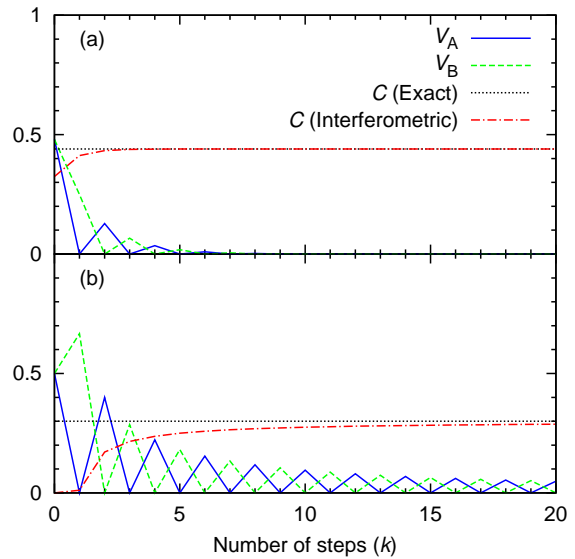}
\caption{ (Color online)
The concurrence $\mathcal{C}^{(k)}(\hat{\rho})$ (dot-dashed lines),
determined from the two-qubit visibilities $\mathcal{V}_\mathrm{AB}$,
and the single-qubit visibilities
$\mathcal{V}^{(k)}_\mathrm{A}$ and $\mathcal{V}^{(k)}_\mathrm{B}$ in
the $k$-th distillation step are shown for (a)
$\hat{\rho}_{\epsilon, \lambda} = \lambda |\phi_\epsilon \rangle \langle \phi_\epsilon |
+ \frac{1-\lambda}{2} ( | \uparrow \downarrow \rangle \langle \uparrow \downarrow | +
| \downarrow \uparrow \rangle \langle \downarrow \uparrow | )$ with $(\epsilon, \lambda) = (0.5,0.8)$
and (b) $\hat{\rho} =
\frac{1}{2}|\uparrow \uparrow \rangle \langle \uparrow \uparrow | +
\frac{2}{5}|\Psi_+ \rangle \langle \Psi_+ | + \frac{1}{10}|\Psi_-
\rangle \langle \Psi_- |$, where
$|\phi_\epsilon \rangle = (\epsilon | \uparrow \uparrow \rangle + | \downarrow \downarrow \rangle)/\sqrt{1+ \epsilon^2}$
 and $|\Psi_\pm \rangle =
\frac{1}{\sqrt{2}}(| \downarrow \uparrow \rangle \pm | \uparrow
\downarrow \rangle)$.
The case (a) is typical, showing rapid
convergence to $\hat{\rho}_\mathrm{dis}$, while (b) is an asymptotic
case with slow convergence. For comparison, the exact value
\protect\cite{Wootters_conc} of $\mathcal{C}(\hat{\rho})$ is given
(dotted lines).} \label{EXAMPLE}
\end{figure}

{\it Examples.}---
In Fig.~\ref{EXAMPLE}, the concurrence is determined at each $k$-th iteration step,
for two examples of $\hat{\rho}$, using $\mathcal{V}_\mathrm{AB}$ and
Eq.~\eqref{Concurrence1}. For typical cases of
$\hat{\rho}$ (non-asymptotic case) [Fig.~\ref{EXAMPLE}(a)],
$\mathcal{V}^{(k)}_{j=\mathrm{A,B}}$ vanishes rapidly within a few steps,
and the determined value $\mathcal{C}^{(k)}(\hat{\rho})$ approaches
to the exact value $\mathcal{C} (\hat{\rho})$ more rapidly.
In this case, the deviation
of $\mathcal{C}^{(k)}(\hat{\rho})$ from $\mathcal{C} (\hat{\rho})$
is estimated \cite{footnote} as $|\mathcal{C} (\hat{\rho})
-\mathcal{C}^{(k)}(\hat{\rho})| \propto (\mathcal{V}^{(k)}_j)^2$
for small
$\mathcal{V}^{(k)}_j$ ($\lesssim 0.1$).
Thus, one
can determine a precise value of $\mathcal{C}$ even before
the complete distillation.
The properties of
particular types of $\hat{\rho}$ are given below.

(i) When $\hat{\rho}$ is pure, only one distillation step is necessary,
since $\mathcal{V}^{(k)}_\mathrm{A} =
\mathcal{V}^{(k)}_\mathrm{B}$ for all $k$ due to
the complementarity \cite{Jakob}.
Note that
$(\mathcal{V}^{(k=0)}_{j=\mathrm{A,B}})^2 + \mathcal{C}^2 (\hat{\rho}) = 1$
for $f_\mathrm{A} = f_\mathrm{B} = 1$.

(ii) When $\hat{\rho}$ is separable and uncorrelated, $\hat{\rho} =
\hat{\rho}_\mathrm{A} \otimes \hat{\rho}_\mathrm{B}$, the local
properties of A and B are independent. Therefore, only two
steps are required, $\mathcal{V}_\mathrm{AB}=0$, and
$\mathcal{C}(\hat{\rho})=0$. Particularly, when either
$\hat{\rho}_\mathrm{A}$ or $\hat{\rho}_\mathrm{B}$ is pure, its
single-qubit visibility is one at $k=0$, and $\hat{\rho}$ cannot be
distilled as $\hat{\rho}_\mathrm{dis}$ vanishes. When $\hat{\rho}$
is separable but has classical correlations, on the other hand, more
than two steps are neccessary, $\mathcal{V}_\mathrm{AB} \ne 0$, and
$\mathcal{C} (\hat{\rho})=0$.

(iii) When $\hat{\rho}$ is a Werner state or a Bell-diagonal state
\cite{Bennet_conc,Werner}, the distillation is not necessary.

(iv) There is the so-called asymptotic case
\cite{Verstraete_filter}, where large number of steps are
necessary [Fig.~\ref{EXAMPLE}(b)] and most
states are abandoned by the filtering ($f_j^{(k \to \infty)} \to
0$).

\begin{table}[bt]
\begin{tabular}{@{} c c c c c @{}}
  \hline \hline
  State & Distillation & Quantification & Total & Tomography \\
  \hline
  Bell & 2400 \;(0) & $9\times100 + 3\times200$ & 3900 & 360 \\
  Werner & 2400 \;(0) & $9\times500 + 3\times4000$ & 18900 & 38700 \\
  $\bf{I}/4$ & 2400 \;(0) & $9\times100 + 3\times7800$ & 26700 & 24300 \\
  $\hat{\rho}'_{0.6}$ & 2400 \; (0) & $9\times300 + 3\times3800$ & 16500 & 30600 \\
  $\hat{\rho}_{0.9, 0.6}$ & 7200 \;(1) & $9\times500 + 3\times6800$ & 32100 & 54900\\
  $\hat{\rho}_{0.5, 0.8}$ & 40800 \;(5) & $9\times600 + 3\times9600$ & 75000 & 38700 \\
  \hline \hline
\end{tabular}
\caption{Monte Carlo simulation \cite{Altepeter} of
the minimum number of individual copies of
a given state $\hat{\rho}$, that
need to be used to determine its concurrence
[or $s_0 (-1 + \lambda_1 + \lambda_2 -q \lambda_3)/2$ in Eq.~\eqref{Concurrence1}]
within $\pm 0.01$ statistical error in our scheme (fourth column) and by full
state tomography \cite{Tomography} (fifth).
In our scheme,
it is the sum of the number of necessary copies for the distillation (second column)
with $k_\mathrm{dis}$ iterative steps, and that for the quantification (third)
consisting of nine measurement settings for the determination of $\vec{v}_{j,l}$
and three settings for the three maxima $\lambda_l$'s; the parentheses show $k_\mathrm{dis}$.
In the distillation, the copies are used to achieve and to check
$\mathcal{V}_j^{(k_\mathrm{dis})} < 0.1$;
for the tested states,
this condition of $\mathcal{V}_j^{(k_\mathrm{dis})}$ is enough
to obtain $\mathcal{C}$
within the $\pm 0.01$ error.
On the other hand,
among available tomography schemes, we consider here the most efficient one
with nine measurement settings and four detectors.
We test six representative states usually tested in entanglement
detection \cite{Altepeter,Kwiat},
a Bell state $|\Psi_0 \rangle = \frac{1}{\sqrt{2}} ( | \uparrow \uparrow \rangle + | \downarrow \downarrow \rangle )$,
a Werner state $\hat{\rho}_W = \frac{2}{3}{|\Psi_0 \rangle \langle \Psi_0 |} + \frac{1}{3}\frac{\bf{I}}{4}$,
$\bf{I}/4$,
$\hat{\rho}'_p = |\Psi_0 \rangle \langle \Psi_0 |
+ \frac{p-1}{2}(|\uparrow \uparrow \rangle \langle \downarrow \downarrow| + |\downarrow \downarrow \rangle \langle \uparrow \uparrow|)$,
and $\hat{\rho}_{\epsilon, \lambda}$ (introduced in Fig.~\protect\ref{EXAMPLE})
with two different sets of $(\epsilon, \lambda)$.
Note that the efficiency of our scheme strongly depends on $\hat{\rho}$
(as the tomography) and becomes worse for states more filtered (those with
smaller $N/M$); $N/M = 1$ (no distillation; $k_\mathrm{dis}=0$) for the first four states,
$0.95$ for $\hat{\rho}_{0.9, 0.6}$, and $0.42$ for $\hat{\rho}_{0.5, 0.8}$.}
\end{table}

Below, we propose an optimal way of determining concurrence.
After the distillation,
one first determines all the rotation vectors $\vec{v}_{j,l}$,
by observing $\langle \delta n_{1 \mathrm{A}} \delta n_{1 \mathrm{B}} \rangle$ in
nine different settings of $\mathrm{BS}_\mathrm{3A}$ and $\mathrm{BS}_\mathrm{3B}$
and comparing the results with the compact form of
$\langle \delta n_{1 \mathrm{A}} \delta n_{1 \mathrm{B}} \rangle$
(derived before). Then, one measures
$\langle \delta n_{1 \mathrm{A}} \delta n_{1 \mathrm{B}} \rangle$ (thus $\lambda_l$)
at the determined setting
of $\mathrm{BS}_{3j=\mathrm{A(B)}}$
at $\vec{v}_{\mathrm{A(B)},l}$.
We emphasize that a {\em crude} determination of
$\vec{v}_{j,l}$ is enough for a precise detection of $\lambda_l$ and $\mathcal{C}$.
It is because $\lambda_l$ is a local maximum, around which
a small error ($\sim \delta$) in the direction
$\vec{v}_{j,l}$ for $\lambda_l$ causes only a much smaller error ($\sim \delta^2$) in
the value of $\lambda_l$. This makes our scheme efficient.
Table I shows that for states not much filtered,
our scheme is as efficient as the tomography \cite{Tomography} for the quantification
of the initial state.
For the distillation and the quantification together, it can be
more efficient than
previous tomographic schemes, e.g., in Ref.~\cite{Wang}; the previous schemes require
roughly 2-4 times larger number of state copies than our scheme, as they require
the tomography twice (once before and once after the distillation).
Moreover, our scheme improves previous distillations~\cite{Kwiat,Wang}, as it is applicable
to unknown states.
Therefore,
our scheme is practically useful, in the situation \cite{Altepeter,Enk}
that for unknown states, the existing schemes are less efficient
than the tomography and virtually require it.

{\it Conclusion.}--- We have proposed a
``quantum
entanglement concentrator'', in which the entanglement of an
arbitrary unknown two-qubit state is distilled and determined.
We remark the following meaningful features.

First, our scheme is within experimental reach and applicable
to generic types of qubits, as it has only
local operations
using a tunable beam splitter, currently
available \cite{Kwiat}.
Second, we show that even for {\em mixed} states,
concurrence and Lorentz singular values are
directly and experimentally
accessible, interestingly from the extrema of two-qubit interference;
concurrence has been determined experimentally
only for a pure state \cite{Walborn}.
This motivates to study
the features of the singular values \cite{Verstraete_Bell}.
Third, entanglement quantification can be closely related with
distillation \cite{Kwiat,Bennett,Pan}. In our scheme, the former
can be done after the latter.
Finally, our scheme may be practically useful (e.g., for teleportation~\cite{Verstraete_tele}),
as it achieves the
distillation and the quantification within one framework.
It would be valuable to generalize our scheme to
larger systems of multiple qubits,
where tomography error estimation
becomes less feasible.

We thank J. B. Altepeter, N. Gisin, Hee Su Park, and Tzu-Chieh Wei for valuable discussions,
and especially the group of P. G. Kwiat for the numerical code for the tomography.
This work was supported by KAIST-HRHRP.


\begin{thebibliography}{99}
\bibitem{footnote} The details of the estimation will be
given elsewhere.
\end{thebibliography}
\end{document}